\documentstyle[psfig,conf_iap,10pt]{article}
\begin{document}

\heading{ The Nature of the Low-redshift Ly$\alpha$ Clouds } 

\par\medskip\noindent
\author{ J. Michael Shull }

\address{Department of Astrophysical \& Planetary Sciences, 
CASA and JILA, \\ 
Campus Box 391, University of Colorado, Boulder CO 80309 (USA)  }

\begin{abstract}
I discuss recent HST observations and interpretation of
low-$z$ Ly$\alpha$ clouds toward nearby Seyferts and QSOs,
including their frequency, space density, estimated mass, association 
with galaxies, and contribution to $\Omega_b$. 
Our HST/GHRS detections of 73 Ly$\alpha$ absorbers 
with N$_{\rm HI} \geq 10^{12.6}$ cm$^{-2}$ along 11 sightlines 
covering pathlength $\Delta(cz) = 86,000$ km~s$^{-1}$
show $f(>N_{\rm HI}) \propto N_{\rm HI}^{-0.6 \pm 0.2}$.  
A group of strong absorbers toward PKS~2155-304 may be 
associated with gas $(400-800)h_{75}^{-1}$ kpc from 4 large galaxies,  
with low metallicity ($\leq0.004$ solar) and D/H 
$\leq 2 \times 10^{-4}$.  At low-$z$, we derive a metagalactic ionizing 
radiation field and Ly$\alpha$-forest baryon density 
$J_0 = (1.1 \pm 0.4) \times 10^{-23}$ ergs cm$^{-2}$ s$^{-1}$
Hz$^{-1}$ sr$^{-1}$ and $\Omega_b = (0.008 \pm 0.004) 
h_{75}^{-1} [J_{-23} N_{14} b_{100}]^{1/2}$ for clouds of
characteristic size $b = (100~{\rm kpc})b_{100}$.
\end{abstract}

\section{Introduction}

Since the discovery of the high-redshift Ly$\alpha$ forest over 
25 years ago, these abundant absorption features in the spectra
of QSOs have been used as evolutionary probes of the intergalactic
medium (IGM), galactic halos, and now large-scale structure
and chemical evolution.  The rapid evolution in the distribution
of lines per unit redshift, $d{\cal N}/dz \propto (1+z)^{\gamma}$
($\gamma \approx 2.5$ for $z \geq 1.6$), was consistent with a 
picture of these
features as highly ionized ``clouds'' whose numbers and sizes were
controlled by the evolution of the IGM pressure, the metagalactic 
ionizing radiation field, and galaxy formation.  
Early observations also suggested that Ly$\alpha$ clouds had 
characteristic sizes $\sim10$ kpc, were much more abundant than 
($L_*$) galaxies, and showed little clustering in velocity space. They were
interpreted as pristine, zero-metallicity gas left over from 
the recombination era.  One therefore expected low-redshift ($z < 1$)
absorption clouds to show only traces of H~I, due to  
photoionization and evaporation in a lower pressure IGM.
All these ideas have now changed with new data.  

One of the delightful spectroscopic surprises from the
{\it Hubble Space Telescope} (HST) was the discovery of  
Ly$\alpha$ absorption lines toward the quasar 3C~273 at 
$z_{\rm em} = 0.158$ by both the Faint Object Spectrograph (FOS, 
(Bahcall et al. 1991) and the Goddard High Resolution Spectrograph 
(GHRS, Morris et al.  1991).  In this review, I will describe (\S2) 
the current status of our group's long-term program with the HST and VLA
to define the parameters and nature of the low-redshift Ly$\alpha$ forest.  
In \S3, I discuss related theoretical work on the metagalactic ionizing 
background, $J_{\nu}(z)$,  and the contribution of low-$z$ 
Ly$\alpha$ clouds to the baryon density, $\Omega_b$.

\section{HST Survey of low-$z$ Ly$\alpha$ Absorbers}

The frequency of low-$z$ Ly$\alpha$ lines reported by the
HST/FOS Key Project, 
$d{\cal N}/dz = (24.3 \pm 6.6)(1+z)^{0.58 \pm 0.50}$
for lines with $W_{\lambda} \geq 320$ m\AA\ (Bahcall et al. 1996),
was considerably higher than a simple extrapolation from the high-redshift
forest.  These higher-N$_{\rm HI}$ absorbers exhibit associations with 
galaxies ($D \leq 200h_{75}^{-1}$ kpc) about half the time 
(Lanzetta et al. 1995).

\begin{figure}
  \centerline{\vbox{
  \psfig{figure=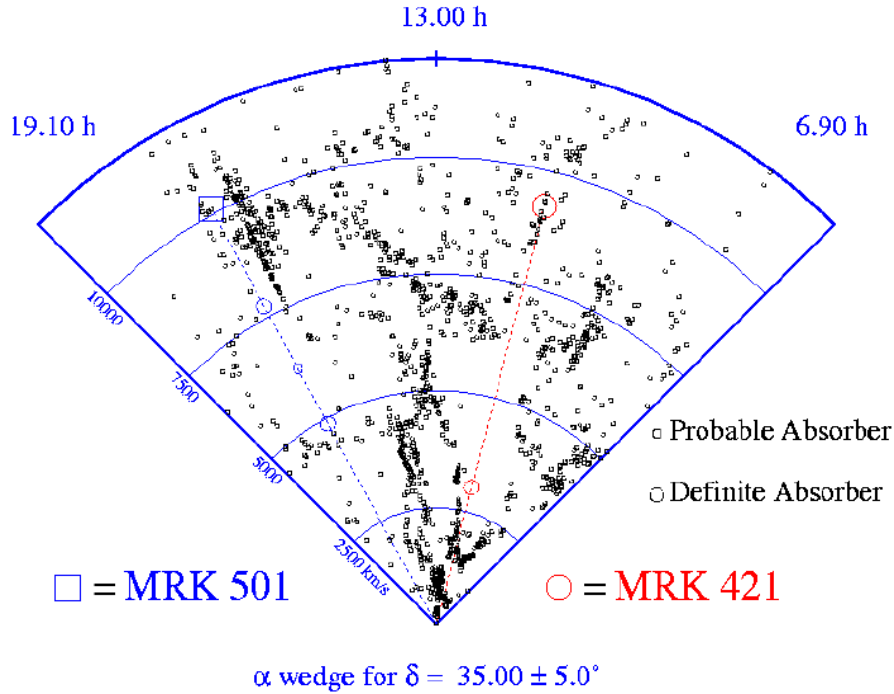,height=10cm} }}
  \caption[]{Pie-diagram distributions of recession velocity and RA 
   of bright (CfA survey) galaxies and four Ly$\alpha$ absorbers 
   toward Mrk~501 and Mrk~421 (Shull, Stocke, \& Penton  1996). Two of 
   these systems lie in voids, with the nearest bright galaxy located 
   more than $4 h_{75}^{-1}$ Mpc from the absorber.  }
  \end{figure}
 
In HST cycles 4--6, our group began GHRS studies of lower-N$_{\rm HI}$
absorbers toward 11 bright targets (Stocke et al. 1995; Shull, Stocke, \& 
Penton  1996). These low-$z$ targets were chosen because of their 
well-mapped distributions of foreground galaxies (superclusters and voids).
Our studies were designed to measure the distribution of Ly$\alpha$
absorbers in redshift ($z \leq 0.08$) and column density 
($12.5 \leq {\rm N}_{\rm HI} \leq 16$), to probe their association
with galaxies, and to measure their clustering and large-scale
structure.  Toward 11 targets, we detected 73 Ly$\alpha$ systems 
over a pathlength $(c \Delta z) \approx 86,000$ km~s$^{-1}$.  In cycle 7,
we will observe 14 more sightlines with the Space Telescope Imaging 
Spectrograph (STIS) to double our Ly$\alpha$ sample.  The locations of 
Ly$\alpha$ absorbers toward two of our first sightlines are shown in Figure 1.  

In our first 4 sightlines,  
the frequency of absorbers with N$_{\rm HI} \geq 10^{13}$ cm$^{-2}$ 
was $\langle d{\cal N}/dz \rangle \approx (90 \pm 20)$,
corresponding to a local space density, $\phi_0 = (0.7~{\rm Mpc}^{-3})
R_{100}^{-2} h_{75}$ for absorber radius $(100~{\rm kpc})R_{100}$.  
This space density is $\sim40$ times that of bright ($L_*$)
galaxies, but similar to that of dwarf galaxies with $L \approx 0.01 L_*$.  
From a statistical, nearest-neighbor analysis, we found that the Ly$\alpha$ 
clouds have some tendency to associate with large structures of galaxies 
and to ``avoid the voids''.  However, for the lower column systems,
the nearest bright galaxies are often too far to be physically associated 
in hydrostatic halos or disks (Maloney 1993; Dove \& Shull 1994).  Of 
10 absorption systems first analyzed (Shull et al. 1996), 3 lie in voids,
with the nearest bright galaxies several Mpc distant.   In several
cases, we identified dwarf H~I galaxies within 100--300 kpc using the VLA
(Van Gorkom et al. 1996). 
 \begin{figure}
    \centerline{\vbox{
    \psfig{figure=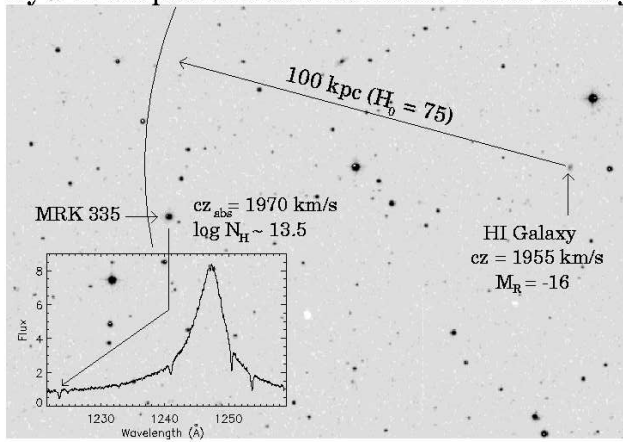,height=7.cm} }}
    \caption[]{Overlay of galaxy field around Mrk~335, showing a 
     dwarf galaxy at 1955 km~s$^{-1}$ at nearly the same redshift
     as the 1970 km~s$^{-1}$ Ly$\alpha$ absorber ($N_{\rm HI} = 
     10^{13.5}$ cm$^{-2}$).  The offset distance is 
     $\sim95  h_{75}^{-1}$ kpc.  }
   \end{figure}
Figure 2 shows one system toward Mrk~335, where a dwarf galaxy with
$M_{\rm HI} \approx (7 \times 10^7~M_{\odot}) h_{75}^{-2}$ and offset
distance $\sim(100~{\rm kpc})h_{75}^{-1}$ is seen
at heliocentric velocity $cz = 1955$~km~s$^{-1}$, remarkably near
to that of the Ly$\alpha$ absorber. Thus, some of the lower-N$_{\rm HI}$
absorbers appear to be associated with dwarf galaxies, although
much better statistics are needed.   

In HST/cycle 6, we observed 7 more sightlines (Penton, Shull, \& Stocke
1998) with the GHRS/G160M.   With better data, we were able
to detect weaker Ly$\alpha$ absorption lines, down to 20 m\AA\
(N$_{\rm HI} = 10^{12.6}$ cm$^{-2}$) in some cases.  Many of the new 
sightlines exhibit considerably more Ly$\alpha$ absorbers;  for these
11 sightlines,  $\langle d{\cal N}/dz \rangle = 250 \pm 40$ 
for N$_{\rm HI} \geq 10^{12.6}$ cm$^{-2}$ or one line
every 1200 km~s$^{-1}$.  Although there is wide variation, 
this frequency is almost 3 times the value (one every 3400
km~s$^{-1}$) reported earlier (Shull et al. 1996) for N$_{\rm HI} 
\geq 10^{13}$ cm$^{-2}$. For a curve of growth with
$b = 25$ km~s$^{-1}$, the 61 Ly$\alpha$ absorbers with 
$12.6 \leq \log {\rm N}_{\rm HI} \leq 14.0$ follow a distribution 
$f(\geq N_{\rm HI}) \propto N_{\rm HI}^{-0.6 \pm 0.2}$.  
These results may change somewhat after correcting 
for incompleteness, line blending, and the GHRS sensitivity function
(Penton et al. 1998). 
\begin{figure}
   \centerline{\vbox{
   \psfig{figure=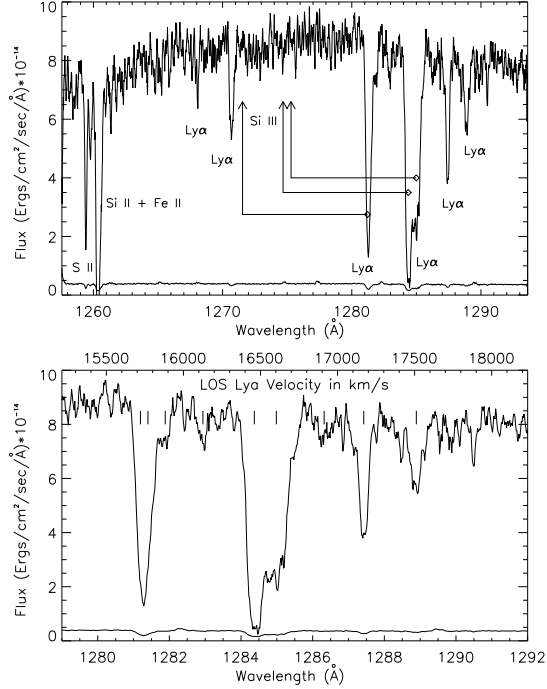,height=10.cm} }}
   \caption[]{ HST/GHRS (G160M) spectrum of PKS~2155-304
    (Shull et al. 1998a) shows multiple
   Ly$\alpha$ absorption systems between 1281--1290 \AA\
   ($cz = 15,700 - 17,500$ km~s$^{-1}$).  Upper limits on
   Si~III $\lambda1206.50$ absorption at 1274.7 \AA\ and 1275.2 \AA\
   correspond to [Si/H] $\leq 0.007$ solar abundance.  Strong lines
   at 1259--1261 \AA\ are Galactic interstellar absorption.   }
\end{figure}

I turn now to the extraordinary sightline toward PKS~2155-304 (Bruhweiler et 
al. 1993).  This target exhibits 12 Ly$\alpha$ absorbers (Fig. 3), 
including a group of strong systems between $cz =$ 15,700 and  
17,500 km~s$^{-1}$.  The strong absorbers have an estimated combined 
column density N$_{\rm HI} = (2-5) \times 10^{16}$ cm$^{-2}$, based on 
Lyman-limit absorption seen by ORFEUS (Appenzeller et al. 1995). 
Using the VLA (Van Gorkom et al. 1996; Shull et al. 
1998a), we have identified these absorbers with the very extended halos 
or intragroup gas associated with four large galaxies at
the same redshift (Fig. 4).  The offsets from the sightline to these
galaxies are enormous. 
\begin{figure}
   \centerline{\vbox{
   \psfig{figure=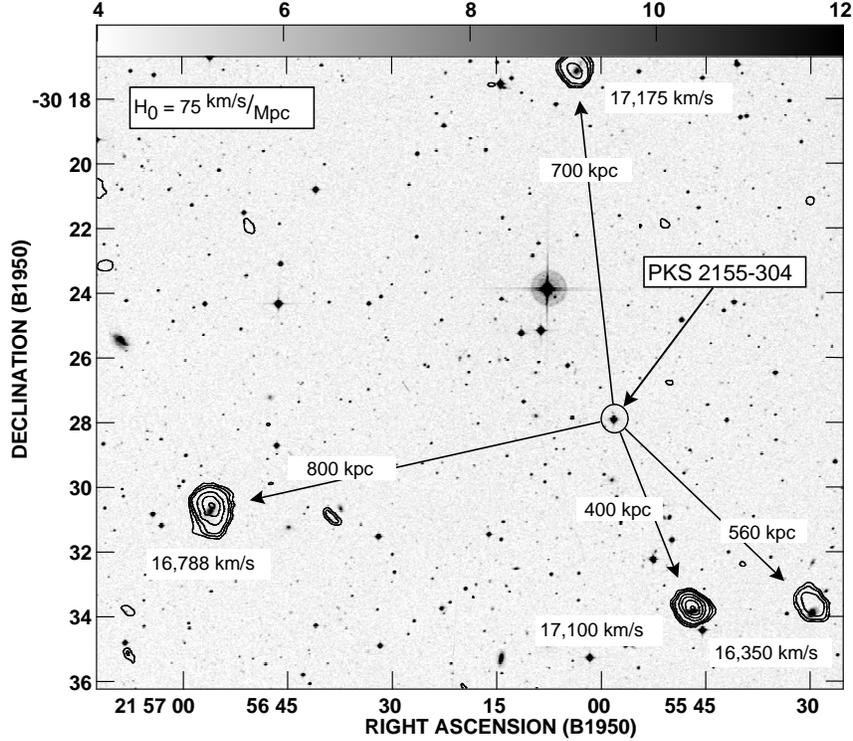,height=10.cm} }}
   \caption[]{VLA field of 21-cm emission toward PKS~2155-304
    at velocities (16,000 -- 17,300 km~s$^{-1}$) near the Ly$\alpha$
    absorbers.  Four H~I galaxies are detected at projected offsets
    of $(400-800)h_{75}^{-1}$ kpc. At least two galaxies, to the south
   and southwest, appear to be kinematically associated with Ly$\alpha$ 
   absorbers at 16,460 and 17,170 km~s$^{-1}$. } 
\end{figure}
Despite the kinematic associations, it would be challenging to make
a dynamical association with such galaxies.   
One must extrapolate from the $10^{20}$ cm$^{-2}$ 
columns seen in galactic 21-cm emission to the range $10^{13-16}$
cm$^{-2}$ probed by Ly$\alpha$ absorption.  Much of the strong Ly$\alpha$ 
absorption may arise in gas of wide extent, $\sim1$ Mpc in diameter, 
spread throughout the group of galaxies at $z = 0.057$.
Assuming $\langle {\rm N}_{\rm HI} \rangle \approx 10^{16}$ cm$^{-2}$ and 
applying corrections for ionization (H$^{\circ}$/H $\approx 3 \times 10^{-4}$ 
for $J_0 = 10^{-23}$ and 600 kpc cloud depth) and for helium mass 
($Y = 0.24$), the total gas mass could be $\sim10^{12}~M_{\odot}$.  

These absorbers offer an excellent opportunity to set  
stringent limits on heavy-element abundances and D/H in low-metallicity gas
in the far regions of such galaxies.  For example, no Si~III $\lambda1206.50$ 
absorption is detected ($W_{\lambda} \leq 23$ m\AA\ or 
N$_{\rm SiIII} \leq 1.0 \times 10^{12}$ cm$^{-2}$ 
at $2\sigma$) at wavelengths corresponding to the strong Ly$\alpha$ 
absorbers at 1281.25 \AA\ and 1284.75 \AA.   Over a range of 
photoionization models for (H$^{\circ}$/H) and (Si$^{+2}$/Si), this limit
corresponds to an abundance (Si/H) $\leq 0.004 ({\rm Si/H})_{\odot}$ 
for an assumed N$_{\rm HI} = 2 \times 10^{16}$ 
cm$^{-2}$ and 300--600 kpc cloud depth (Shull et al. 1998a).    
The lack of observed C~IV $\lambda1549$ absorption leads to
similar limits.  A rudimentary analysis of the lack of 
observed D~I (Ly$\alpha$) absorption in the blueward wings of the 
strong H~I lines suggests that (D/H) $\leq 2 \times 10^{-4}$.  These limits 
can be improved with more sophisticated profile fitting and future 
data from HST/STIS and FUSE.    

The H~I toward PKS~2155-304 appears to represent gas with the 
lowest detected metallicity.  Was this gas was once inside the galaxies 
at $cz = 17,000 \pm 1000$ km~s$^{-1}$, or is it pristine? 
We can perhaps answer this question by deeper spectral 
searches for traces of metals.   The origin of the lower-column Ly$\alpha$ 
systems would seem to be more diverse, possibly arising in extended 
halos or debris disks of dwarf galaxies, large galaxies, and small 
groups (Morris \& van den Bergh 1994).

\section{Theoretical Implications}

A primary theoretical issue is whether low-$z$ clouds 
have any relation to the evolution of the baryons in the 
high-$z$ forest.  A quick estimate suggests that the low-$z$
absorbers could contain a substantial (25\%) fraction  
of the total baryons estimated from Big Bang nucleosynthesis,
$\Omega_{\rm BBN} = (0.036 \pm 0.007) h_{75}^{-2}$ (Burles \& Tytler 1997).   
Consider those Ly$\alpha$ systems with N$_{\rm HI} \geq 10^{13}$
cm$^{-2}$, for which one can derive the space density $\phi_0$,  
\begin{equation}
    \frac {d{\cal N}} {dz} = \phi_0 (\pi R_0^2) \frac {c}{H_0} 
            \approx 100  \; . 
\end{equation}  
The major uncertainty in deriving absorber masses is the ionization
correction, which depends on the profile of gas density around 
the cloud centers.  Assume, for simplicity, that
$n_H(r) = n_0 (r/r_0)^{-2}$, and adopt photoionization equilibrium
at rate $\Gamma_{\rm HI}$ at 20,000~K.  The ionizing radiation field 
is $J_{\nu} = J_0 (\nu / \nu_0)^{-\alpha_s}$ with 
$\alpha_s \approx 1.8$ and 
$J_0 = (10^{-23}~{\rm ergs~cm}^{-2}~{\rm s}^{-1}~{\rm Hz}^{-1}~
{\rm sr}^{-1}) J_{-23}$.  The H~I column density integrated
through the cloud at impact parameter $b$ is, 
\begin{equation}
   {\rm N}_{\rm HI}(b) = \frac { \pi n_0^2 r_0^4 \alpha_H^{(A)}  
       (1 + 2n_{\rm He}/n_{\rm H}) } {2 \Gamma_{\rm HI} b^3 } \; .
\end{equation}
We can solve for $n_0r_0^2$ and find the total gas mass within
$b = (100~{\rm kpc}) b_{100}$ for a  
fiducial column density N$_{\rm HI} = (10^{14}~{\rm cm}^{-2}) N_{14}$, 
\begin{equation}
   M_{\rm cl}(b) = [4 \pi n_0 r_0^2 b (1.22 m_H)] =   
     (1.6 \times 10^{9}~M_{\odot}) N_{14}^{1/2}
     J_{-23}^{1/2} b_{100}^{5/2} \; , 
\end{equation}
which yields a cloud closure parameter in baryons,
\begin{equation} 
\Omega_b \approx \phi_0(b) M_{\rm cl}(b) = (0.008 \pm 0.004) 
           J_{-23}^{1/2} N_{14}^{1/2}
           b_{100}^{1/2} h_{75}^{-1} \;. 
\end{equation} 
Note that $\Omega_b$ is insensitive to the scaling parameters:  
\begin{equation}
    \Omega_b \propto [J_{-23} N_{14} b_{100}]^{1/2}     
          \left[ \frac {T_e}{20,000~{\rm K}} \right] ^{0.38}
          \left[ \frac {4.8}{\alpha_s + 3} \right] ^{1/2}
           h_{75}^{-1}  \; .
\end{equation}
For the spherical-cloud model, the radiation field,
cloud size, and column-density distribution probably each contribute
30--40\% to the uncertainty in $\Omega_b$, while temperature $T_e$ and
ionizing spectral index $\alpha_s$ contribute 10\%, for an overall 
uncertainty of 50\%.  However, as with the high-$z$ forest, the greatest 
uncertainty in $\Omega_b$ lies in the
cloud geometry and radial profile.  These parameters can only be
understood by building up statistics through many sightlines,  
particularly multiple targets that probe the same cloud structures.

We have also increased our understanding of the metagalactic ionizing
background radiation and the ``Gunn-Peterson'' opacities,
$\tau_{\rm HI}(z)$ and $\tau_{\rm HeII}(z)$.
Using a new cosmological radiative transfer code and IGM opacity model,
Fardal, Giroux, \& Shull (1997) model the ionization fractions of 
H~I and He~II in a fluctuating radiation field due to quasars and 
starburst galaxies.  In this work, we have calculated the metagalactic 
ionizing radiation field, $J_{\nu}(z)$, using QSO and stellar emissivities
and including cloud diffuse emission and new (somewhat lower) IGM opacities
derived from {\it Keck} Ly$\alpha$ forest spectra. 

\begin{figure}
   \centerline{\vbox{
   \psfig{figure=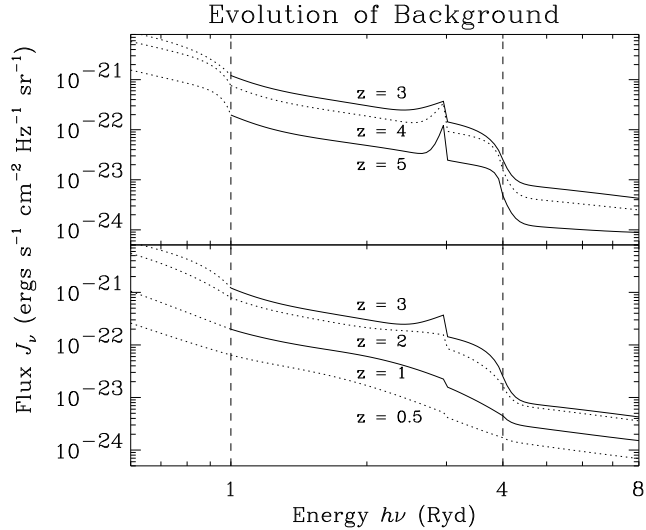,height=7cm} }}
   \caption[]{ Spectrum, $J_{\nu}(z)$, of ionizing background
   from redshift $z = 5 \rightarrow 0.5$ from new opacity and radiative
   transfer model (Fardal, Giroux, \& Shull 1997).  }
\end{figure}
 
Figure 5 illustrates the evolution of $J_{\nu}$ from $z = 5 \rightarrow 0.5$,
peaking at $z \approx 3$.  At $z < 2$, the absorption breaks 
at 1 and 4 Ryd become much less prominent and $J_{\nu}$ drops rapidly. 
At low redshift ($z < 0.5$), $J_{\nu}$ depends
both on the local (Seyfert) luminosity function and on the opacity
model. David Roberts, Mark Giroux, and I have recalculated
these quantities (Shull et al. 1998b) using extrapolated EUV emissivities of
low-redshift Seyferts from our IUE-AGN database
(Penton \& Shull, unpublished) and a new low-$z$ opacity model.
We find $J_0 = (1.1 \pm 0.4) \times 10^{-23}$ 
ergs cm$^{-2}$ s$^{-1}$ Hz$^{-1}$ sr$^{-1}$ at $z = 0$, very close to 
our adopted scaling parameter, $J_{-23} = 1$.  

We clearly still have an enormous amount to learn
about the nature and distribution of the low-redshift Ly$\alpha$
clouds.  It seems likely that future studies
may uncover valuable information about their connection to
large-scale structure and to the processes of galaxy formation
and evolution.  

%   \begin{center}
%   \begin{tabular}{l r c}
%   \multicolumn{3}{l}{{\bf Table 1.} Here is an example for Tables } \\
%  % % \hline
%  \\
%  \multicolumn{1}{c}{$\lambda_{\rm helio}$}&\multicolumn{1}{c}{$w_{\rm obs}$}&
%  \multicolumn{1}{c}{Identification}\\
%  \multicolumn{1}{c}{(\AA)}&\multicolumn{1}{c}{(\AA)}&
%  \multicolumn{1}{c}{ }\\
%  \hline
%  \\
%    1215.7 & 12.0  & Ly$\alpha$ \\
%    3333.3 & 7.1   & H$_2$      \\
%    3332.8 & 0.7   & D$_2$      \\
%    9797.1 & 0.1   & Pas s\^ur  \\
%   \\
%   \hline
%  \end{tabular}
%  \end{center}
%

\acknowledgements{
Our Ly$\alpha$ observations were made with the NASA/ESA {\it Hubble Space
Telescope} supported by grant GO-06586.01-95A through the Space Telescope 
Science Institute. Theoretical work was supported by NSF grant AST96-17073.
My colleagues in observational work include John Stocke
and Steve Penton (Colorado), Jacqueline Van Gorkom (Columbia
University), and Chris Carilli (NRAO).  Theoretical work was done 
in collaboration with Mark Giroux and Mark Fardal (Colorado) and
undergraduate research student David Roberts (Cornell).
}

\vfill

\begin{iapbib}{99}{

\bibitem{App95} Appenzeller, I., Mandel, H., Krautter, J., Bowyer, S.,
   Hurwitz, M., Grewing, M., Kramer, G., \& Kappelmann, N.  1995,
   \apj, 439, L33

\bibitem{Bah91} Bahcall, J. N., Januzzi, B. T., Schneider, D. P.,
   Hartig, G. F., Bohlin, R., \& Junkkarinen, V. 1991, \apj, 377, L5

\bibitem{Bah96} Bahcall, J. N., et al. 1996, \apj, 457, 19 

\bibitem{Bru93} Bruhweiler, F. C., Boggess, A., Norman, D. J., Grady, C. A., 
    Urry, C. M., \& Kondo, Y. 1993, \apj, 409, 199

\bibitem{Bur97} Burles, S., \& Tytler, D.  1997, AJ, submitted

\bibitem{Dov94} Dove, J. B., \& Shull, J. M. 1994, \apj, 423, 196

\bibitem{Far97} Fardal, M. A., Giroux, M. L., \& Shull, J. M. 1997, 
      \aj, submitted

\bibitem{Lan95} Lanzetta, K. M., Bowen, D. V., Tytler, D., \& Webb, J.  K.
   1995, \apj, 442, 538

\bibitem{Mal93} Maloney, P. 1993, \apj, 414, 41

\bibitem{Mo95} Morris, S. L., \& van den Bergh, S.  1994, \apj, 427, 696 

\bibitem{Mor91} Morris, S. L., Weymann, R. J., Savage, B. D., \&
   Gilliland, R. L. 1991, \apj, 377, L21

\bibitem{Mor93} Morris, S. L., et al. 1995, \apj, 419, 524 

\bibitem{Pen98} Penton, S., Shull, J. M., \& Stocke, J. T. 1998,
    in preparation 

\bibitem{Sto95} Stocke, J. T., Shull, J. M., Penton, S., Donahue, M.,
   \& Carilli, C. 1995, \apj, 451, 24

\bibitem{Shu96} Shull, J. M., Stocke, J. T., \& Penton, S. 1996, \aj, 111, 72 

\bibitem{Shu98a} Shull, J. M., Penton, S., Stocke, J. T., Giroux, M. L.,
   Van Gorkom, J. H., \& Carilli, C.  1998a, \aj, in preparation 
 
\bibitem{Shu98b} Shull, J. M., Roberts, D., Giroux, M. L., Penton, S., \&
    Fardal, M. A. 1998b, \aj, in preparation 

\bibitem {Van96} Van Gorkom, J. H., Carilli, C. L., Stocke, J. T., 
    Perlman, E. S., \& Shull, J. M.  1996, \aj, 112, 1397 

}
\end{iapbib}
\end{document}